\documentstyle[12pt,aaspp4,psfig]{article}
 
\lefthead{T. Heckman et al.}
\righthead{Absorption-Lines in Galactic Winds}
 
\slugcomment{.}
 
\begin{document}

\title{The Detection of the Diffuse Interstellar Bands in 
Dusty Starburst Galaxies}

\author{Timothy M. Heckman$^1$}
\affil{Department of Physics and Astronomy, Johns Hopkins
University, Homewood Campus, 3400 North Charles Street, Baltimore, MD 21218}
\author{Matthew D. Lehnert$^1$}
\affil{Max-Plank-Institut f\"ur extraterrestrische Physik, Postfach 1603,
D-85740 Garching, Germany}

\parindent=0em
\vspace{4cm}
 
1. Visiting astronomers, Kitt Peak National Observatory and Cerro Tololo
Interamerican Observatory, NOAO,
operated by AURA, Inc. under cooperative agreement with the
National Science Foundation.

\newpage
\parindent=2em

\begin{abstract}
We report the detection of the Diffuse Interstellar Bands (``$DIBs$'') in
the optical spectra of seven far-infrared-selected starburst galaxies.
The $\lambda$6283.9 \AA\ and $\lambda$ 5780.5 \AA\
features are detected
with equivalent widths of $\sim$ 0.4 to 1 \AA\ and 0.1 to 0.6 \AA\ respectively.
In the two starbursts with the highest quality spectra
(M82 and NGC2146), four other weaker $DIBs$ at 
$\lambda$ 5797.0 \AA, 6010.1 \AA, 6203.1 \AA,
and 6613.6 \AA\
are detected with equivalent widths of $\sim$ 0.1 \AA. The region over
which the $DIBs$ can be detected ranges from $\sim$ 1 kpc in the less
powerful starbursts, to several kpc in the more powerful ones.
The gas producing the $DIBs$ is more kinematically quiescent on-average
than the gas producing the strongly-blueshifted $NaI\lambda\lambda$5890,5896
absorption in the same starbursts. We show that the $DIBs$ in these intense
starbursts are remarkably similar to those in our Galaxy: the relative
strengths of the features detected are similar, and the equivalent widths
follow the same dependence as Galactic $DIBs$ on $E(B-V)$ and $NaI$ column
density. While the ISM in starbursts is heated by a photon and cosmic
ray bath that is $\sim$ 10$^3$ times more intense than in the diffuse
ISM of the Milky Way, the gas densities and pressures are also
correspondingly larger in starbursts. This ``homology'' may help
explain the strikingly similar $DIB$ properties.
\end{abstract}

\keywords{
Galaxies: Starburst -- Galaxies: Nuclei -- Galaxies: ISM -- ISM: Molecules
-- ISM: Dust,Extinction -- Line: Identification}

\newpage

\section{Introduction}

The Diffuse Interstellar Bands (``$DIBs$'') have been studied for over 60 years,
since Merrill (1934) first established their origin in the interstellar medium.
Despite decades of intensive investigation, the identity of the carrier or
carriers of
the $DIBs$ has not been established (see the comprehensive
review by Herbig 1995). The 
most likely candidates are large
carbon-rich molecules (e.g. Sonnentrucker et al 1997), perhaps
Polycyclic Aromatic Hydrocarbons (PAH's - Salama et al 1999).
The strongest and best-studied $DIBs$ in the optical spectrum are known
empirically to trace the $HI$ phase of the ISM, with
strengths that correlate well with the line-of-sight color excess $E(B-V)$,
the $HI$ column density, and the $NaI$ column density as probed with the
$NaI \lambda\lambda$5890,5896 (``$NaD$'') doublet (see Herbig 1993).

To date, $DIBs$ have been observed almost exclusively in our own Galaxy,
and to a limited extent in the Magellanic Clouds (Morgan 1987).
Supernova 1986G in NGC 5128 (Cen A) allowed the detection of
$DIBs$ produced within the famous dusty gas disk in this elliptical
galaxy (di Serego Alighieri \& Ponz 1987). Most recently, 
Gallagher \& Smith (1999) have reported the possible discovery
of the $DIB$ at $\lambda$6283.9 \AA\ in the spectra of two
``super starclusters'' near the
nucleus of the prototypical starburst galaxy M 82. This is intriguing,
since it suggests that the $DIB$ carriers are present at a normal level
even in the ISM of an intense starburst, in which the ambient radiation
intensity and gas pressure are orders-of-magnitude higher than
in the diffuse gas in the Milky Way disk (e.g. Colbert et al 1999).

We have recently analyzed the properties of the interstellar $NaD$ absorption-
line in the spectra of 18 high-luminosity, infrared-selected (dusty) starbursts
(Heckman et al. 2000 - hereafter HLSA).
In the course of this analysis, we examined the 7 starbursts with
the highest quality spectra for the presence of $DIBs$ (section 2). As we report
in section 3 below,
we have detected one or both of the
$\lambda$6283.9 \AA\ and 
$\lambda$5780.5 \AA\ $DIB$ features (normally the two strongest $DIBs$ in the
optical spectral region) in all seven cases.
We have also been able
to map the spatial distribution of the $DIBs$. 
These data allow us to directly compare the properties
of the $DIBs$ in the ISM of
these extreme starbursts to sight-lines in the Galaxy having similar
gas column density and reddening (section 4).

\section{Observations \& Data Analysis}

Details concerning the following are given in HLSA, so we only summarize
the most salient points here.

The starburst sample presented here is a subset of the 32 galaxies observed
by HLSA. The HLSA sample itself was selected from two far-infrared-bright
samples: the Armus, Heckman, \& Miley (1989) sample of galaxies
with very warm far-IR colors and the Lehnert \& Heckman (1995) sample
of far-IR-bright disk galaxies seen at high inclination. The combined
sample is representative of the far-IR-galaxy phenomenon, but is not complete.

HLSA found that the $NaD$ line was of predominantly interstellar origin
in 18 of the 32 galaxies, while cool stars contributed significantly
to the line in the other 14 cases. 
The plethora of weak
absorption features in the spectra of cool stars greatly complicate
the detection of the $DIBs$, while the strength of $DIBs$
in our Galaxy correlate strongly with the ISM $NaI$ column density. Thus,
the galaxies in the present paper were drawn exclusively from the
18 ``interstellar-dominated'' objects in HLSA.
We then selected the objects in HLSA having the
highest signal-to-noise spectra obtained with
a resolution better than $\sim$ 100 km s$^{-1}$
(see below). This results in a sample of 7 galaxies, as listed in
Table 1.

The observations were undertaken in 1993 and 1994
using two different facilities: the 4-meter Blanco Telescope with the
Cassegrain Spectrograph at $CTIO$ and the 4-meter Mayall Telescope with the
RC Spectrograph at $KPNO$. The spectral resolution ranged from 1.1 \AA\
FWHM in the $KPNO$ data to 1.8 \AA\ FWHM in the $CTIO$ data. Details regarding
spectrograph configurations
are listed in Table 2 of HLSA. 

The spectra were all processed using the standard {\it LONGSLIT} package
in $IRAF$ (bias-subtracted, flat-fielded using spectra of a quartz-lamp,
geometrically-rectified and wavelength-calibrated using a $HeNeAr$ arc lamp,
and then sky-subtracted). See HLSA for details. No explicit correction was
made for the presence of weak telluric absorption-features, but these
are not a problem for our analysis. The strongest feature of relevance
is the O$_2$ band from $\sim$ 6276 to 6284 \AA\ (e.g. Figure 2a in
Benvenuti \& Porceddu 1989). Fortunately,
the redshifts
of our galaxies are sufficient to move the $\lambda$6283.9 \AA\ $DIB$ out
from under this feature.

The spectra were analyzed  using the interactive {\it SPLOT} spectral fitting
package in $IRAF$. In all cases, a one-dimensional ``nuclear'' spectrum was
extracted, covering a region with a size set by the slit width and summed over
5 pixels
in the spatial direction (the resulting aperture is typically 2 by 4 arcsec).
The
corresponding linear size of the projected aperture is generally a few hundred
parsecs to a few kpc in these galaxies (median diameter 600 pc).
This is a reasonable match to the typical sizes of powerful starbursts
like these (e.g. Meurer et al 1997; Lehnert \& Heckman 1996). Prior
to further analysis, each 1-D spectrum was normalized to unit intensity by
fitting it with, and then dividing it by, a low-order polynomial. 
Similar one-dimensional spectra for off-
nuclear regions were extracted over the spatial region with adequate signal-
to-noise in the continuum for each galaxy.

It is essential to remove the myriad absorption features due to cool stars
from the spectra before searching for the relatively weak $DIB$ features. We 
have therefore used the average spectrum of several Galactic K giant stars as a 
template. After redshifting the normalized stellar template to the galaxy
rest-frame,
we have iteratively scaled and subtracted the template from the normalized
galaxy spectrum until the residuals in the difference spectrum were minimized
in the spectral regions that exclude potentially detectable interstellar
features. 
The scale factors found for the stellar template
imply that cool stars typically contribute 20 to 30\% of the continuum light
at $\sim$ 6000\AA. This is consistent with both the less rigorous estimates
reported
in HLSA for these galaxies, and with theoretical
expectations for red supergiants in a mature metal-rich starburst
(Bruzual \& Charlot 1993; Leitherer et al 1999). 
To compensate
for the effects of the continuum subtraction,
we added back an equivalent amount of featureless continuum.
Thus, the depths and equivalent widths of the $DIBs$ in the original
data are preserved by our analysis.
As an example, we show the spectrum of the nucleus of M82 before and after
the subtraction of a suitably-scaled K-star spectrum in Figure 1.
These final processed spectra are shown in Figures 2 and 3.

We have estimated the uncertainties in our measurements in two ways.
First, we compared 
the measurements for the four galaxies in the sample for which we have more
than one independent spectrum (taken at a different position angle).
Second, we have calculated the rms noise in the cool-star-subtracted spectra
and used this to calculate the implied uncertainties (assuming standard
error propogation for Poissonian noise). We report these uncertainties
in Tables 1 through 3.

\section{Results}

\subsection{The Identified Features}

The two most conspicuous $DIBs$ along typical sight-lines in the ISM of
our Galaxy are the strong, relatively narrow features at
$\lambda$6283.9 \AA\ and
$\lambda$5780.5 \AA\ (Herbig 1995). In all but one case, both features are
within the
wavelength coverage of our spectra. The next strongest $DIBs$ in the Milky
Way in the relevant spectral region are at 5797.0 \AA, 6010.1 \AA, 6203.1 \AA,
and 6613.6 \AA. We have searched for all these features in our spectra.

We turn our attention first to the $\lambda$6283.9 \AA\ feature, which is 
the strongest feature in the Milky Way, and is not seriously confused by
stellar photospheric lines in the starburst spectra.
As can be seen in Figure 2, the $\lambda$6283.9 $DIB$ is detected in 6 of the
7 starburst nuclei (Table 1).
The only exception is NGC 6240, where the feature would
lie within the blue shoulder of the very strong and broad [OI]$\lambda$6300
nebular emission-line, making it very difficult to detect. 
In the other six cases, the equivalent
width of this $DIB$ ranges from $\sim$ 0.4 to 0.9 \AA\, with a normalized
residual intensity at line-center of 0.83 to 0.94.
These values correspond to some of 
the strongest features seen along sight-lines in the ISM of the Milky Way
(e.g. Chlewicki et al 1986; Benvenuti \& Porceddu 1989).

Weaker absorption due to the $\lambda$5780.5 $DIB$
is definitely present in three of the seven members of our sample 
(NGC2146, M82, and NGC6240), and possibly present in three more
(NGC1614, NGC1808, and NGC3256).
No measurement can be made
in IRAS10565+2448, since 
the feature lies just outside our spectral passband. In the five cases
in which both features are detected,
the $\lambda$5780.5 $DIB$ is typically about 25\% as strong as
the $\lambda$6283.9 feature, compared to a mean value of about 45\%
along comparably-reddened lines-of-sight in the Milky Way
(Chlewicki et al 1986; Benvenuti \& Porceddu 1989). 
We emphasize that the measurement 
of the
$\lambda$5780.5 $DIB$ is difficult in our spectra owing to its proximity to the 
comparably
strong
stellar photospheric CrI+CuI$\lambda$5782 feature (with which it is badly
blended). We estimate that this introduces an uncertainty
of $\pm$ 50 m\AA\ in the quoted equivalent widths (which is generally larger
than the formal measurement uncertainties estimated above).

The two starburst nuclei with the strongest $\lambda$6283.9 \AA\ $DIB$ feature
are NGC 2146 and M82. These two spectra also have the
highest signal-to-noise and (along with IRAS10565+2448) have the best
spectral resolution and broadest wavelength coverage in our sample.
In these two spectra, several other weaker $DIB$ features can be identified,
namely those at 5797.0 \AA, 6010.1 \AA, 6203.1 \AA,
and 6613.6 \AA\ (Figure 3).
The equivalent widths of these features are $\sim$100
m\AA\, or typically 10 to 15\% as large as those of the
$\lambda$6283.9 \AA\ feature. These relative strengths agree reasonably well
with Galactic $DIBs$ (Chlewicki et al 1986; Benvenuti \& Porceddu 1989;
Herbig 1995).
We summarize this information in Table 2, and note that similarly-weak
features could be present in the noisier spectra of the other five
members of our sample.

\subsection{Kinematics}

We have measured the width and centroid of the $\lambda$6283.9 \AA\ $DIB$
 feature
in all cases but NGC 6240 (where we have instead used the $\lambda$5780.5 
\AA\ $DIB$).
The measured line widths (Table 3) range from $\sim$ 5 to 9 \AA. The intrinsic
width of the $\lambda$6283.9 ($\lambda$5780.5) $DIB$ in the Milky Way is $\sim$
4 (2) \AA\ (Herbig 1995).
Taking our instrumental resolution into account,
the implied Doppler broadening of the $DIBs$ due to macroscopic motions
in the starburst ISM ranges from FWHM 160 to 430 km s$^{-1}$. In
four of the seven cases,
these Doppler widths are smaller than the widths of
the $NaD$ doublet (by 25 to 60\%). In NGC1808, NGC2146 and M82, the $DIB$
$NaD$ lines have the roughly the same Doppler widths.
Interestingly, HLSA find that these are the three cases in the present sample in
which the nuclear $NaD$ lines do not show significant blueshifts with respect
to the galaxy systemic velocity ($v_{sys}$). 

The centroids of the $DIBs$ are within $\sim$ 100 km
s$^{-1}$ of $v_{sys}$. However, in all four cases with strongly
blueshifted
$NaD$ lines (NGC1614, NGC3256, IRAS10565+2448, and NGC6240), the $DIBs$
are mildly blueshifted (by $\sim$ 50 to 110 km s$^{-1}$) with velocities
that are intermediate between $v_{NaD}$ and $v_{sys}$. The velocities
of the $DIB$ and $NaD$ absorbers roughly agree with one another 
(and lie close to $v_{sys}$) in the other three
cases. This kinematic information is summarized in Table 3.

Taken together, these results suggest that 
the $DIBs$ trace gas that is more quiescent on-average than that probed by
the $NaD$ line. That is, the $NaD$ absorption in the  four ``outflow''
nuclei is probably produced by
a combination of quiescent material ($v \sim v_{sys}$
with smaller Doppler width) and disturbed,
outflowing material. 
The bulk of the $DIB$ absorption would be associated with the former,
and this component would dominate both the $NaD$ and $DIB$ absorption
in the three other cases in our sample.

\subsection{Spatial Extent}

We have used our long-slit data to map the 
extra-nuclear spatial extent of the $DIBs$
in these galaxies.
As listed in Table 1, these sizes
range from
$\sim$ 1 to 6 kpc. The absorbing region is larger
(3 to 6 kpc) in the more powerful starbursts (NGC1614, NGC3256,
IRAS10565+2448, and NGC6240, with $logL_{bol}$ = 11.3 to 12.0 $L_{\odot}$),
and smaller (0.9 to 1.8 kpc) in the less powerful cases
(NGC1808, NGC2146, and M82, with $logL_{bol}$ = 10.5 to 10.7 $L_{\odot}$).
In the nearby (less powerful) starbursts, these sizes reflect the
extent of the absorbing material. In the more distant (more powerful)
starbursts, these sizes are lower limits set by the
region with adequate signal-to-noise
in the stellar continuum.

\section{Discussion}

\subsection{Comparison to Galactic DIBs}

The strengths of the prominent Galactic $DIBs$ correlate well
with the column densities of both $HI$ and $NaI$ 
and with the reddening
along the line-of-sight
(e.g. Chlewicki et al 1986; Herbig 1993).
This implies
that the $DIB$ carrier is most plausibly associated with the cool
atomic phase of the ISM. We can use the data discussed in HLSA
to estimate the values for $N_{NaI}$ and $E(B-V)$ in our sample
of seven starbursts, to see if the $DIBs$ in our starburst sample
obey the same empirical relations
defined by the ISM of the Milky Way.

We follow HLSA and derive estimates for $N_{NaI}$ using the average
of the values obtained from 
the classical ``doublet ratio'' method (Spitzer 1968) and the
variant described by Hammann et al (1997). We estimate the line-of-sight
reddening to the stellar continuum using 
the observed colors
compared to 
theoretical models for a starburst stellar population
(Leitherer et al 1999; see HLSA for details). We list
the results in Table 1.

Our best-measured $DIB$ by-far is the strong $\lambda$6283.9 feature.
The data compiled by Chlewicki et al (1986) and Benvenuti \& Porceddu (1989)
show that the mean ratio of the equivalent width of this feature
and the color excess is $<W_{6284}/E(B-V)>$ = 1.2 \AA\ for heavily-reddened
Galactic sight-lines. For our small starburst sample we find a similar result:
$<W_{6284}/E(B-V)>$ = 0.8 \AA. This is shown
in Figure 4 where we have plotted
$W_{6284}$ {\it vs.} $E(B-V)$ for a large sample of Galactic sight-lines
using the extensive data compled by Herbig (1993). To compare our starburst data
directly to this Galactic data we have converted the values given by
Herbig (1993) for the equivalent width of the $\lambda$5780.5
\AA\ $DIB$ into estimated values for $W_{6284}$ assuming that the mean
ratio measured by Chlewicki et al (1986) and Benvenuti \& Porceddu (1989)
applies ($W_{6284}/W_{5780}$ = 2.2). In Figure 5 we have likewise
plotted $W_{6284}$ {\it vs.} $N_{NaI}$ for both the Galactic data
and our starburst data. The starbursts lie at the high-end of 
the relationship defined by $DIBs$ in the
Milky Way. 

\subsection{Relationship to the $\lambda$2175 \AA\ Dust Feature}

Over the years, there has been considerable speculation as
to a possible connection between the $DIBs$ and the strong and broad
feature at $\lambda$2175 \AA\ in the Galactic extinction curve
(see Benvenuti \& Porceddu 1989). In this context, the detection
of strong $DIBs$ in starburst spectra is noteworthy. As shown
by Calzetti et al (1994), the $\lambda$2175 feature is extremely
(undetectably) weak in the UV spectra of starbursts. This
implies that the carriers of the $DIBs$ and the $\lambda$2175 feature
must be quite distinct (in agreement with the conclusions
of Benvenuti \& Porceddu (1989) for the Galactic ISM).

\subsection{Speculations}

On the face of it, the above results may seem surprising given
the extreme differences between the physical
conditions in the ISM of intense starbursts and our own Galactic disk.
The strong starbursts in our sample have bolometric surface
brightnesses of $\Sigma_{bol} \sim 10^{10}$ to $10^{11}$ L$_{\odot}$ 
kpc$^{-2}$
(e.g. Meurer et al. 1997), typical star-formation rates per unit area of
$\Sigma_{SFR} \sim$ 10 M$_{\odot}$ year$^{-1}$ kpc$^{-2}$, and surface
mass densities in gas and stars of
$\Sigma_{gas} \sim \Sigma_{stars} \sim$ 10$^9$ M$_{\odot}$ kpc$^{-2}$ (e.g.
Kennicutt 1998).
These
are roughly 10$^3$ ($\Sigma_{SFR}$), 10$^2$ ($\Sigma_{gas}$) and
10$^1$ ($\Sigma_{stars}$) times larger than the
corresponding values in the disks of normal galaxies.
These values for $\Sigma_{bol}$ 
correspond to a radiant energy density inside the star-forming region that
is roughly 10$^3$ times the value in the ISM of the Milky Way (and see
Colbert et al 1999 for direct measurements of this quantity). 
The rate of mechanical energy deposition
(supernova heating) per unit volume in these starbursts is of-order 10$^3$
times
higher than in the ISM of our Galaxy (e.g. Heckman, Armus, \& Miley 1990),
as is the cosmic ray heating rate (Suchkov, Allen, \& Heckman 1993).
Finally,
simple considerations of hydrostatic equilibrium imply correspondingly
high pressures in the ISM: $P \sim G \Sigma_g
\Sigma_{tot} \sim$ few $\times$ 10$^{-9}$ dyne cm$^{-2}$ (P/k $\sim$
few $\times$ 10$^7$ K cm$^{-3}$, or several thousand times the value
in the local ISM in the Milky Way). These high pressures have been
confirmed observationally (e.g. Heckman, Armus, \& Miley 1990;
Colbert et al 1999).

The interesting result of the above is that despite the extreme conditions
prevailing inside these starbursts,
the dimensionless ratio of the
ISM pressure
to the energy density in UV photons (or cosmic rays) is quite similar
in starbursts and the disk of the Milky Way. This would in turn imply
that (for a given ISM temperature) the ratio of the number densities
of the gas particles and UV
photons (or cosmic rays) would also be similar to their values in the
local ISM. Wang, Heckman, \& Lehnert (1998) have discussed the evidence
that this analysis is correct for the diffuse ionized medium in starbursts
and the disks of normal late-type galaxies. 

This ``homologous'' behavior of the ISM in regions spanning over
three orders-of-magnitude in heating and cooling rates per particle
may help to explain why the ratio of the column
density of $DIB$ carriers to that of both $Na$ atoms (Figure 5) and dust grains 
(Figure 4) appears so similar in extreme starbursts and the ISM of our
own Galaxy. In the absence of a well-understood origin for the $DIBs$,
further speculation seems premature.

\section{Summary}
Despite over six decades of investigation, the nature and origin of
the Diffuse Interstellar Bands remain a mystery (Herbig 1995). We have
presented evidence that - far from being a possibly pathological
property of the local ISM in our Galaxy -  $DIBs$ are probably ubiquitous
in the spectra of far-infrared-bright (dusty) starbursts.

In our own Galaxy, the two most conspicuous $DIBs$ are the features
at $\lambda$6283.9 \AA\ and $\lambda$5780.5 \AA. 
We have detected one or both of these two $DIBs$ in all seven starbursts selected
on the basis of strong interstellar $NaI\lambda\lambda$5790,5796 ($NaD$)
absorption from the larger starburst sample studied by Heckman et
al (2000 - HLSA). The equivalent widths of these features are
$\sim$ 400 to 900 m\AA\ and $\sim$ 100 to 400 m\AA\ for the
$\lambda$6280.9 and $\lambda$5780.5 features respectively. These roughly
correspond to the greatest $DIB$
 strengths observed in the Milky Way
(Herbig 1993;
Chlewicki et al 1986).
In two members of our sample (M82 and NGC2146) the spectra are of
high enough signal-to-noise to detect four other weaker $DIBs$
(at 5797.0 \AA, 6010.1 \AA, 6203.1 \AA, 
and 6613.6 \AA). These have typical equivalent widths of $\sim$ 100 m\AA.
The relative strengths of these $DIBs$ are rather similar to those
in the Milky Way (Herbig 1995; Chlewicki et al 1986;
Benvenuti \& Porceddu 1989).

The $DIBs$ can be mapped over an extensive region in and around the
nuclear starbursts. In the moderately powerful starbursts
($L_{bol}$ = few $\times$ 10$^{10}$ L$_{\odot}$), this region is
$\sim$ 1 kpc in size {\it vs.} several kpc in the more powerful
starbursts ($L_{bol}$ = few $\times$ 10$^{11}$ L$_{\odot}$). The kinematics
of the gas producing the $DIBs$ is evidently more quiescent than that
producing the $NaD$ absorption studied by HLSA. In the four starbursts
with broad and strongly blueshifted $NaD$ lines, the $DIBs$ are less
Doppler-broadened and much less blueshifted ($v_{DIB}$ - $v_{sys}$
$\sim$ -100 km s$^{-1}$).

In the Milky Way, the $DIBs$ are known to trace a dusty atomic phase
of the ISM, since their equivalent widths correlate strongly with
the $HI$ column density, the $NaI$ column density, and the reddening
parameter $E(B-V)$ (Herbig 1995 and references therein). We show that
these starburst $DIBs$ obey the same trends with $N_{NaI}$ and
$E(B-V)$ (e.g. $W_{6284} \sim$ 1.2 $E(B-V)$ \AA\ at log$N_{NaI}
\sim$ 14 cm$^{-2}$). Thus, the abundance of the $DIB$ carrier(s) relative
to $Na$ atoms and dust grains appears to be very similar in intense
starbursts and the diffuse ISM of our own Galaxy.

This seems surprising, given the thousand-fold greater energy density in 
photons and cosmic rays in the ISM of an intense starburst
(e.g. Colbert et al 1999; Suchkov, Allen, \& Heckman 1993). However,
the gas pressures and densities in the starburst ISM are correspondingly
larger as well (e.g. Heckman, Armus, \& Miley 1990). Thus, such
key dimensionless ratios as gas/photon density and
gas-pressure/radiant-energy-density are similar in the ISM of starbursts
and the disks of normal spiral galaxies (Wang, Heckman, \& Lehnert
1998). This apparent ``homology'' may help explain the strikingly
similar $DIB$ properties.

Finally, we point out that starbursts apparently produce strong $DIBs$
without producing a detectable $\lambda$2175 \AA\ dust feature
in their UV spectra (Calzetti et al 1994). This underscores the
quite distinct origin of the two types of features.

\acknowledgements

We thank David Neufeld, Ken Sembach, and Don York for useful conversations
at various stages of this project. The partial support
of this project by NASA grant NAGW-3138 is acknowledged.

\begin{deluxetable}{lcccccc}
\tablecolumns{7}
\tablewidth{0pt}
\tablenum{1}
\tablecaption{Basic Properties}
\tablehead{
\colhead{Galaxy}&\colhead{$W_{6284}$}&\colhead{$W_{5780}$}&
\colhead{Ang. Size}&\colhead{Size}&\colhead{$E(B-V)$}&\colhead{log$N_{NaI}$} \\
\colhead{(1)}&\colhead{(2)}&
\colhead{(3)}&\colhead{(4)}&
\colhead{(5)}&\colhead{(6)}&
\colhead{(7)}}
\startdata
NGC1614        & 940$\pm$40 & 100: & 11 & 3.5 & 0.9 & 14.1 \nl
NGC1808        & 550$\pm$40 & 140: & 17 & 1.2 & 0.9 & 14.3 \nl
NGC2146        & 940$\pm$30 & 360 & 28 & 1.8 & 0.9 & 13.8 \nl
M82            & 880$\pm$30 & 240 & 55 & 0.9 & 1.0 & 13.9 \nl
NGC3256        & 370$\pm$40 & 100: & 18 & 3.0 & 0.6 & 13.8 \nl
IRAS10565+2448 & 530$\pm$90 & ... &  5 & 4.5 & 0.7 & 14.0 \nl
NGC6240        & ... & 640$\pm$70 & 12 & 6.0 & 1.2 & 13.9 \nl
\enddata
\end{deluxetable}

Note. Col. (2) The equivalent width of the $\lambda$6283.9 $DIB$
in m\AA.
Col. (3) The equivalent width of the $\lambda$5780.5 $DIB$
in m\AA. The uncertainty is due primarily
to the accuracy with which contamination by the stellar photospheric
CrI+CuI$\lambda$5782 can be removed. We estimate this leads to an uncertainty
of $\pm$50m\AA. The detection of this $DIB$ is therefore only tentative
in NGC1614, NGC1808, and NGC3256 (indicated by a colon).
Col. (4) The angular size (in arcsec) over which the $\lambda$6283.9 $DIB$
is detectable (the $\lambda$5780.5 $DIB$ was used in NGC6240).
Col. (5) The corresponding physical size (in kpc), for our adopted
$H_0$ = 70 km s$^{-1}$ Mpc$^{-1}$.
Col. (6) The estimated color excess along the line-of-sight to the
stellar continuum, based on the observed continuum color and a model
starburst spectral energy distribution (Leitherer et al 1999; see HLSA
for details).
Col. (7) The logarithm of the estimated column density of $NaI$ atoms
(cm$^{-2}$). These were derived using the standard doublet ratio technique
(Spitzer 1968) and its variant in Hammann et al (1997). See HLSA for
details. Based on an intercomparison of the values obtained by different
techniques, we estimate the uncertainty to be $\pm$0.2 dex.

\begin{deluxetable}{lccccccc}
\tablecolumns{8}
\tablewidth{0pt}
\tablenum{2}
\tablecaption{Weaker DIBs}
\tablehead{
\colhead{Galaxy}&\colhead{$W_{5780}$}&\colhead{$W_{5797}$}&
\colhead{$W_{6010}$}&\colhead{$W_{6203}$}&\colhead{$W_{6283}$}&
\colhead{$W_{6613}$}&\colhead{$\Delta$}\\
\colhead{(1)}&\colhead{(2)}&
\colhead{(3)}&\colhead{(4)}&
\colhead{(5)}&\colhead{(6)}&
\colhead{(7)}&\colhead{(8)}}
\startdata
NGC2146 & 360 & 100 & 130 & 160 & 910 & 120 & 30 \nl
M82     & 240 &  70 & 110 & 120 & 880 & 120 & 30 \nl
\enddata
\tablecomments{
See Figure 2. All equivalent widths are given in m\AA.
Col. (8) Uncertainties in m\AA.}
\end{deluxetable}

\begin{deluxetable}{lccccc}
\tablecolumns{6}
\tablewidth{0pt}
\tablenum{3}
\tablecaption{Kinematic Properties}
\tablehead{
\colhead{Galaxy}&\colhead{$\Delta$$v_{DIB}$}&\colhead{$\Delta$$v_{NaD}$}&
\colhead{$v_{sys}$}&\colhead{$v_{DIB}$}&\colhead{$v_{NaD}$} \\
\colhead{(1)}&\colhead{(2)}&
\colhead{(3)}&\colhead{(4)}&
\colhead{(5)}&\colhead{(6)}}
\startdata
NGC1614        & 300(7.7$\pm$0.6) & 420 &  4760 &  4657 &  4636 \nl
NGC1808        & 240(6.7$\pm$0.6) & 300 &  1001 &  1040 &  1013 \nl
NGC2146        & 160(5.3$\pm$0.4) & 140 &   916 &   932 &   930 \nl
M82            & 160(5.3$\pm$0.4) & 170 &   214 &   268 &   204 \nl
NGC3256        & 220(6.4$\pm$0.6) & 550 &  2801 &  2755 &  2489 \nl
IRAS10565+2448 & 370(9.3$\pm$1.0) & 500 & 12923 & 12840 & 12717 \nl
NGC6240        & 430(8.9$\pm$1.0) & 600 &  7339 &  7232 &  7049 \nl
\enddata
\end{deluxetable}

Note. Col. (2) The Doppler broadening (full-width-at-half maximum) in
km s$^{-1}$ for the $\lambda$6283.9 $DIB$ (the $\lambda$5780.5 $DIB$ was used in
NGC6240). These widths have been corrected for the intrinsic width
of the DIB feature (see text) and for the instrumental resolution
of the spectrograph (see HLSA). The raw, measured line widths
and their associated uncertainties (in \AA) are given in parantheses.
Col. (3) The full-width-at-half-maximum in km s$^{-1}$ of the members of the
$NaI\lambda\lambda$5890,5896 doublet ($NaD$). Uncertainties are $\pm$20 km
s$^{-1}$. Taken from HLSA.
Col. (4) The heliocentric galaxy systemic velocity.
Approximate uncertainties range
from $\pm$10 km s$^{-1}$ for NGC1808, NGC2146, and M82 to
$\pm$50 km s$^{-1}$ for NGC1614 and NGC3256, to $\pm$100 km s$^{-1}$
for NGC6240 and IRAS10565+2448.
See HLSA and references therein.
Col. (5) The heliocentric velocity of the $\lambda$6283.9 $DIB$
(the $\lambda$5780.5 $DIB$ was used in NGC6240). The measurement
uncertainties are $\pm$30 km s$^{-1}$ for NGC2146 and M82,
$\pm$50 km s$^{-1}$ for NGC1614, NGC1808, and NGC3256, and
$\pm$80 km s$^{-1}$ for IRAS10565+2448 and NGC6240. These do not
include any uncertainties in the true value of the rest wavelength for
the $DIB$.
Col. (6) The heliocentric velocity of the $NaD$ doublet taken
from HLSA. Uncertainties are $\pm$20 km s$^{-1}$.

\newpage

\figcaption []
{The spectrum of the nucleus of M82 before (top) and after (bottom)
the subtraction of the scaled spectrum of K giant star. This
subtraction removes the stellar photospheric absorption features
whose presence complicates the detection and measurement of
the $DIBs$ in our sample of starbursts. See text for details.}

\figcaption []
{Spectra of the $DIB$ at $\lambda_{rest}$ =
6283.9 \AA\ in six starburst nuclei and of the $DIB$ at $\lambda_{rest}$ =
5780.5 \AA\ in NGC6240 (denoted by tick marks). These spectra have been
normalised to unit intensity, cleaned of photospheric absorption-lines
due to cool stars by subtraction of a suitably-normalized spectrum
of a K giant star, and then diluted by the addition of featureless
continuum equal in strength to the subtracted starlight. See Figure 1 for
an example of the effect of K-star subtraction on the spectrum of M82.
The absorption feature at $\lambda_{observed} \sim$ 
6278 \AA\ in NGC1808, NGC2146, and M82 is telluric O$_2$. See
text for details.}

\figcaption []
{Spectra of the nuclei of NGC2146 and M82 showing other
$DIBs$. The spectra have been processed as
described in Figure 2 (see text).
The detected features are indicated by
five tick marks denoting the $DIBs$ at $\lambda_{rest}$ = 5780.5 \AA\,
5797.0 \AA, 6010.1 \AA, 6203.1 \AA, and 6283.9 \AA.
The unmarked absorption feature at $\lambda_{observed} \sim$ 
6278 \AA\  is telluric O$_2$. 
See Table 2.}

\figcaption []
{The equivalent width of the Diffuse Interstellar Band at
$\lambda_{rest}$ =6283.9 \AA\ (in m\AA) is plotted {\it vs.} the
color excess $E(B-V)$ for a large sample of Galactic stars
(hollow points) and our starburst nuclei (larger solid points).
The Galactic data come from Benvenuti \& Porceddu (1989),
Chlewicki et al (1986), and Herbig (1993). Since Herbig gave
only the equivalent widths for the weaker $DIB$ at $\lambda$5780.5 \AA\,
we have converted these values to $W_{6284}$ assuming the mean
ratio measured by Benvenuti \& Porceddu (1989) and 
Chlewicki et al (1986): $<W_{6284}/W_{5780}>$ = 2.2. We have
done likewise for NGC6240 in which the $\lambda$ 6283.9
feature is buried under a strong and broad [OI]$\lambda$6300
nebular emission-line (see Figure 3). Note that the starburst
nuclei lie along the relationship defined by the Galactic sight-lines.}

\figcaption []
{The logarithm of the equivalent width of the Diffuse Interstellar Band at
$\lambda_{rest}$ =6283.9 \AA\ (in m\AA) is plotted {\it vs.} the
logarithm of the $NaI$ column density or a large sample of Galactic stars
(hollow points) and our starburst nuclei (larger solid points).
The Galactic data come from Herbig (1993). Since Herbig gave
only the equivalent widths for the weaker $DIB$ at $\lambda$5780.5 \AA\,
we have converted these values to $W_{6284}$ assuming the mean
ratio measured by Benvenuti \& Porceddu (1989) and 
Chlewicki et al (1986): $<W_{6284}/W_{5780}>$ = 2.2. We have
done likewise for NGC6240 in which the $\lambda$ 6283.9
feature is buried under a strong and broad [OI]$\lambda$6300
nebular emission-line (see Figure 3). Note that the starburst
nuclei lie at the upper end of the 
relationship defined by the Galactic sight-lines.}

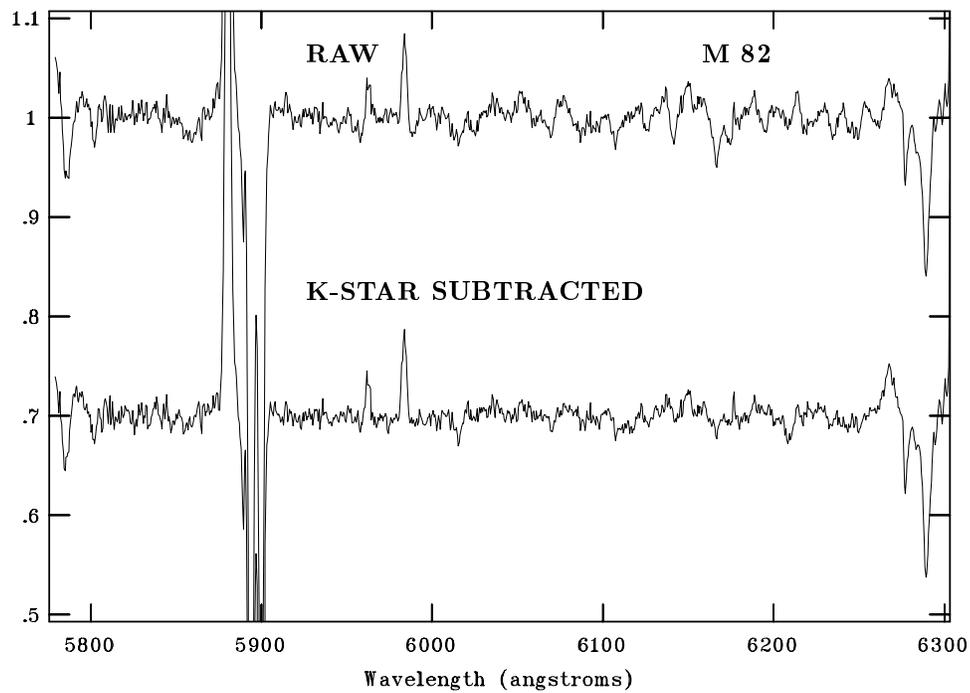
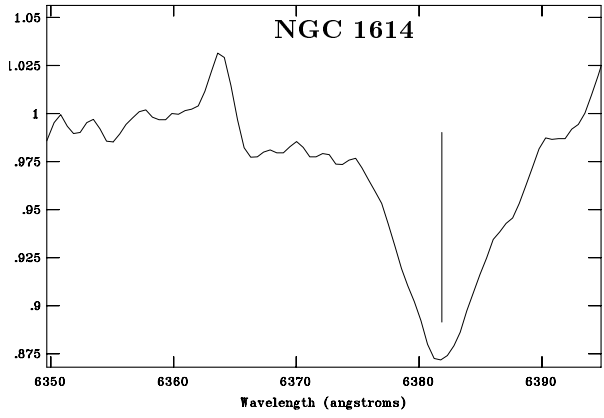
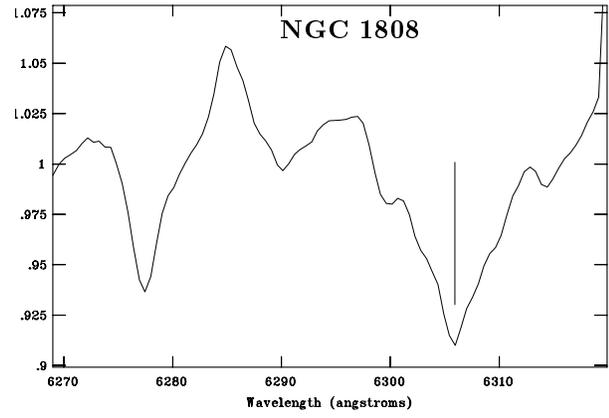
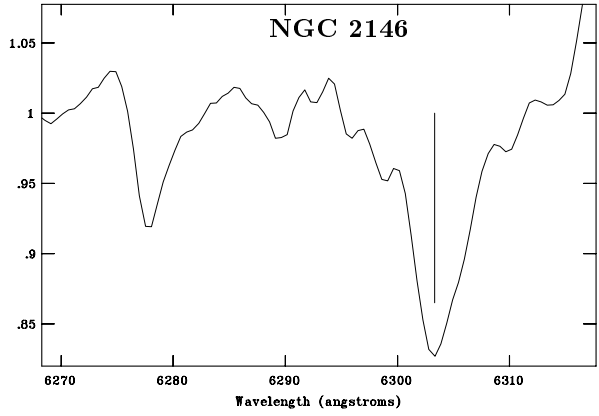
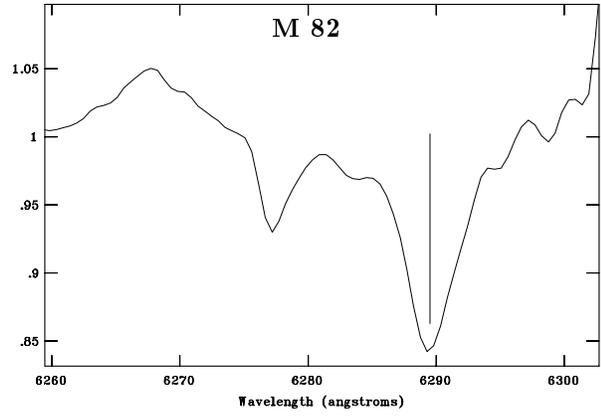
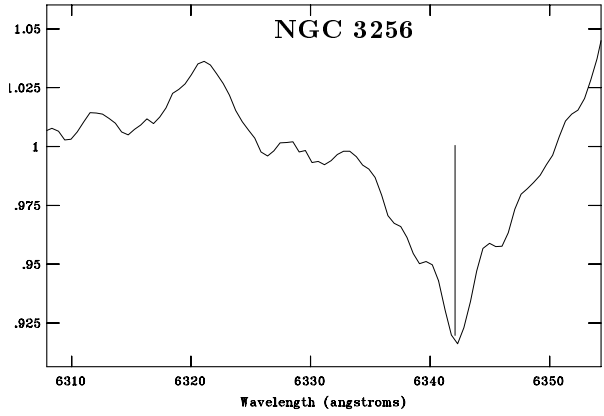
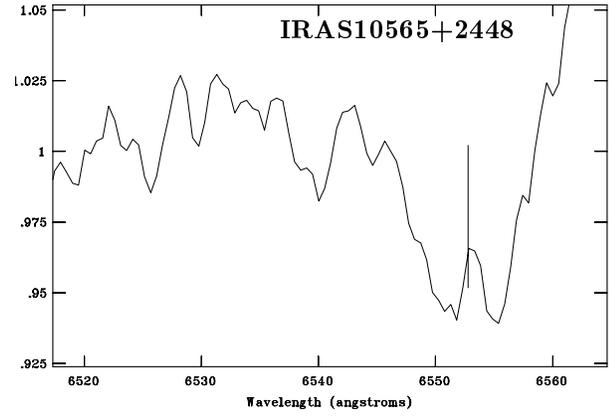
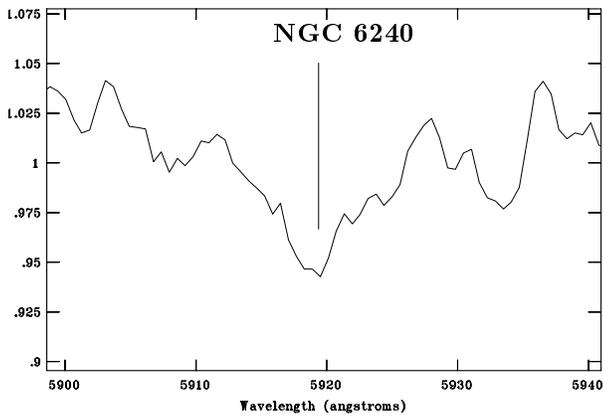
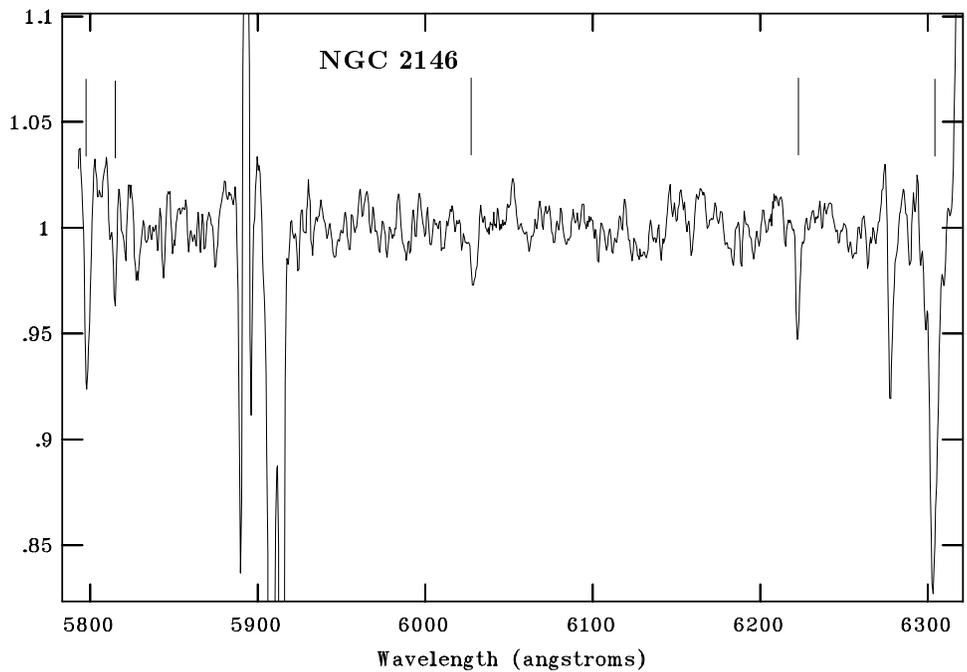
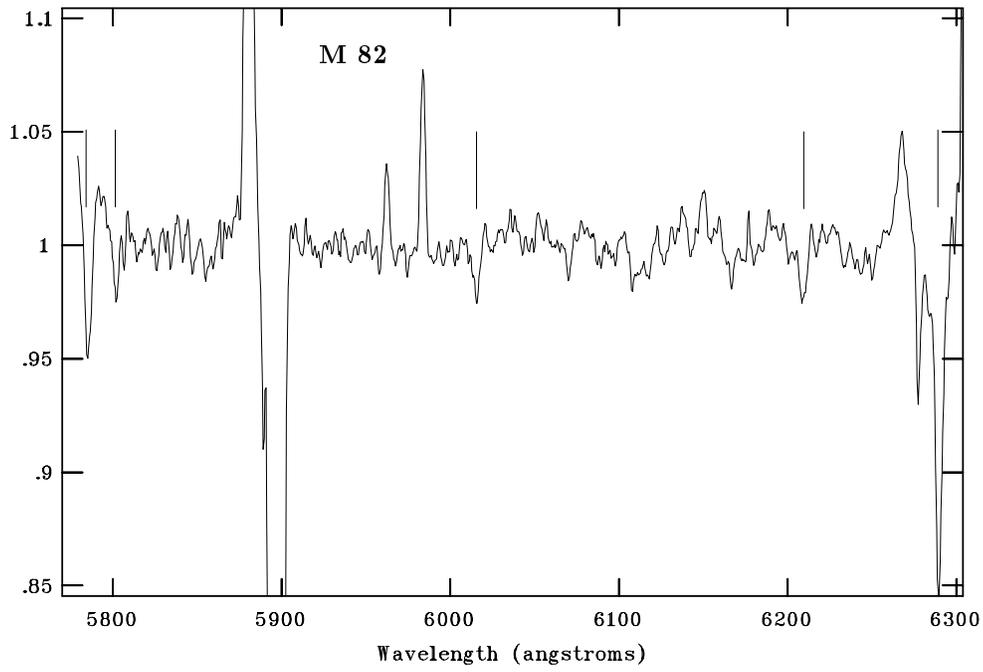
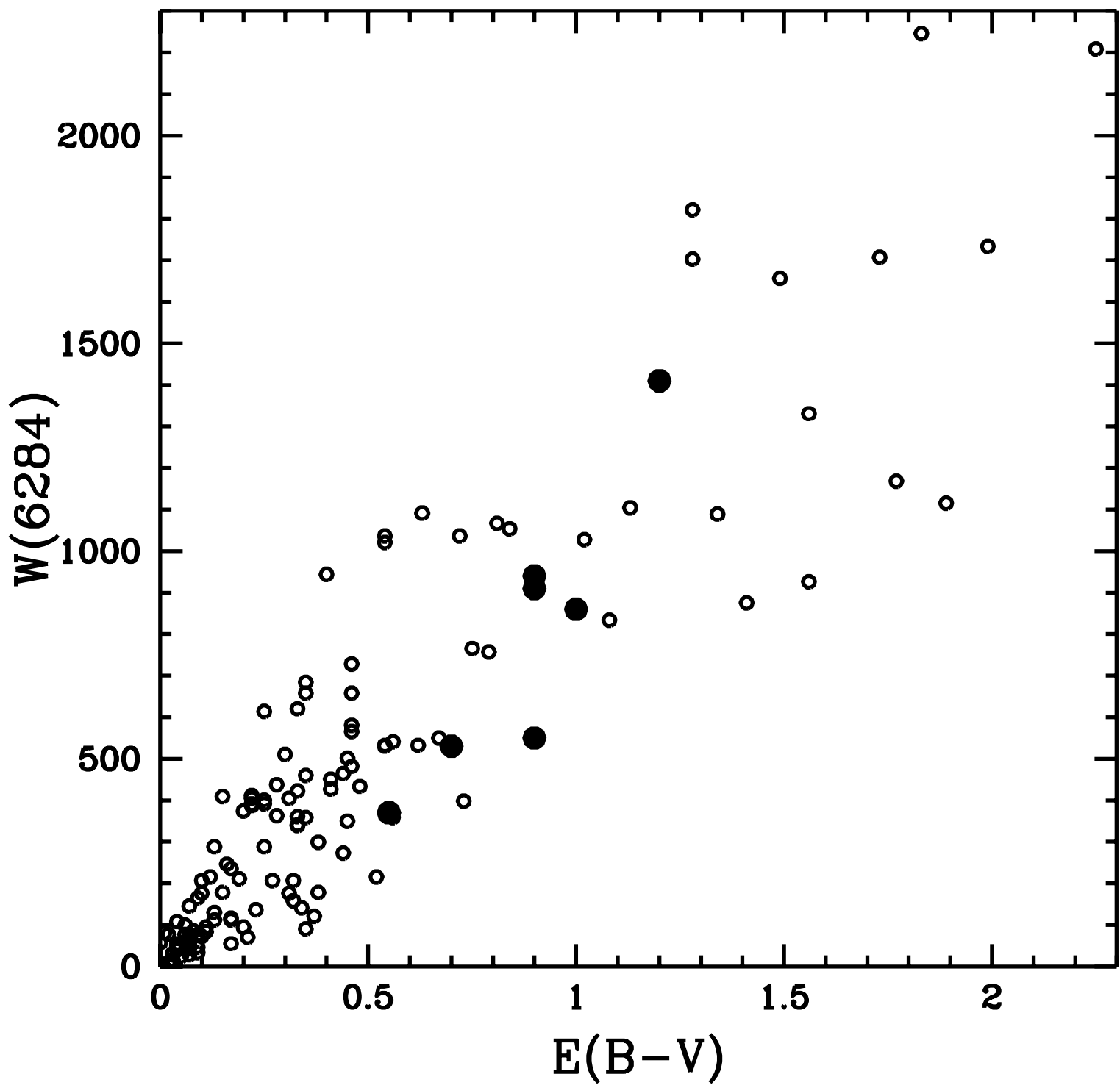
\end{document}